# Losing a Gold Mine?

**Syed Abul Basher**[1]                **Salim Rashid**[2]                **Mohammad Riad Uddin**[3]

**Abstract.** Four rounds of surveys of slum dwellers in Dhaka city during the 2020-21 COVID-19 pandemic raise questions about whether the slum dwellers possess some form of immunity to the *effects* of COVID-19? If the working poor of Bangladesh are practically immune to COVID-19, why has this question not been more actively investigated? We shed light on some explanations for these pandemic questions and draw attention to the role of intellectual elites and public policy, suggesting modifications needed for pandemic research.

**JEL Codes:** I18.

**Keywords:** COVID-19, Immunity, Public health, Ethical approval.

## I. The Pandemic Questions

Suppose the working poor of Bangladesh were somehow immune to COVID-19, would this not be a gold mine for the country? Not only could everyday production continue in both farms and factories but also foreign employers could be persuaded to hire Bangladeshis in droves. Saudi Arabia, Qatar, the UAE, Malaysia, and even China could each be persuaded to hire Bangladeshi workers in the thousands. It was very much in their self-interest to hire workers who could live in their own isolated housing areas, keep production ongoing with minimal risk and save the economy. Both for the domestic economy and for overseas workers it was a unique opportunity. What a gold mine! How was it missed?[4]

Consider the impact on the Bangladesh economy alone. The absolute numbers are eye-catching. Growth in 2020 should have been $324 billion if the pattern of the past is extrapolated, but was $312 billion.[5] The loss in the *constant* dollar is approximately $12 billion. Assuming the additional loss over the next 18 months to be small by comparison, say $3 billion, we get a total loss of $15 billion. A special COVID tax of 5% would generate $750 million. At $1 million per health complex, $0.5 million for infrastructure and $0.5 million for operation, this builds 495 new public health complexes, or one for each Upazila (subdistrict). It still leaves us with $255 million more for public health units to be built in the cities, where the stadiums (and even suitable school playgrounds) should be converted for public health use in pandemics. With such actions, long-run pandemic preparations could have been made for the entire countryside and emergency plans for the cities could have begun. We have not even tried to put a number on the educational loss suffered, which is itself probably minor compared to the psychological harm due to prolonged loss of family income and isolation. Every crisis is an opportunity; was the opportunity taken? It is perhaps time to assess what the intellectual elite of Bangladesh should have done.

---

[1] **Corresponding Author:** Department of Economics, East West University, Jahurul Islam City, Plot A-2, Aftabnagar, Dhaka 1212, Bangladesh. Phone: +8809666775577 ext. 198. Email: syed.basher@ewubd.edu

[2] Center for Urban Studies and Sustainable Development, East West University, Jahurul Islam City, Plot A-2, Aftabnagar, Dhaka 1212, Bangladesh. Email: srashid@ewubd.edu

[3] International Food Policy Research Institute, House 10A, Road 35, Gulshan 2, Dhaka 1212, Bangladesh. E-mail: r.uddin@cgiar.org

[4] The idea of COVID immunity as a powerful employment tool during pandemics has been a topic of ongoing research (see, e.g., Eichenberger et al. 2020, Aranzales et al. 2021).

[5] Authors calculations based on data from BBS (2022).

'Suppose the working poor of Bangladesh were somehow immune to COVID-19', this was our initial premise. One needs a factual background for such strong claims. In so doing, we conducted surveys of slums in Dhaka, matched against timelines of COVID-19 and Eid festivals. As Eid is a time when social distancing is an impossibility, and seems to have been quite ignored, Eid days are times when super spreading events occur. The surveys were deliberately conducted before and after Eid whenever possible, so as to capture the maximal impact of COVID.[6] In total we took four rounds of surveys to track the extent of COVID-19 infection among slum dwellers of Dhaka.[7] Table 1 presents a summary of the findings of the surveys.

**Table 1. Number of households reported COVID-19 infection in their family**

|  | Round 1 (13-25 July, 2020) | Round 2 (26-29 Sept., 2020) | Round 3 (3-8 May, 2021) | Round 4 (27 June - 2 July, 2021) |
|---|---|---|---|---|
| No. of infection | 0 | 0 | 13[i] | 4[ii] |
| No. of households | 255[iii] | 247[iv] | 281[v] | 275[iii] |

i. 1 from village and 12 from slum areas.
ii. 1 from village and 3 from slum areas.
iii. All slum households.
iv. 66 village households and 181 slum households.
v. 97 village households and 184 slum households.

The major question that we asked in the surveys was: 'Do Bangladeshi slum dwellers possess some form of immunity to the effects of COVID-19?' We have not seen any papers that asked this exploratory question. Let us emphasize that when we say 'no study', we mean 'no study that has been publicized to the extent warranted by a pandemic'. If others have tried to make the points we are making, that will be both welcome and important. *Public policy is known for its focus upon orders of magnitude.* This is the crucial point: have the questions been raised and investigated with the emphasis and persistence that such a momentous issue required? In all rounds, we asked the following questions:

- **Round 1:** Is there anyone infected from her/his household?
- **Round 2:** After Eid, is there anyone infected from her/his household?
- **Round 3:** In last 2 months, is there anyone had/has the syndrome or been infected by COVID-19.
- **Round 4:** After Eid-ul-Fitr, is there anyone had/has the syndrome or is infected by COVID-19.

[6] Over our sample period, there were four Eid festivals (Eid ul-Fitr on May 25, 2020; Eid al-Adha on August 1, 2020; Eid ul-Fitr on May 14, 2021; and Eid al-Adha on July 21, 2021).

[7] The timeline of these surveys are: First round (13-25 July, 2020); Second round (26-29 September, 2020); Third round (3-8 May, 2021); and Fourth round (27 June to 2 July, 2021). See Basher et al. (2021) for further details.

Assuming that our slum dwellers are expected to have first-hand knowledge about their near neighbors, during our surveys we also asked questions relating to ten adjoining habitations. Therefore, our findings have potential implications for 2000-2500 slum households.

Let the main points in question be stated briefly, based on a public health/economics perspective:

- If COVID is not debilitating or fatal, then being contagious alone is a public health issue and not a national emergency.
- Since close living quarters like Dhaka slums seem to be potent for spreading covid, why have there been so very few cases in Dhaka slums?
- This low virulence holds even after Eid, which should be a definite 'super spreading' event.

Probable causes, moving from deeper ones to those more apparent:

- Genetic endowments. It was said that Bangladesh has 63% of a gene associated with COVID-19, whereas the Europeans have less than 10% (Zeberg and Pääbo, 2020). What if this very genetic endowment is actually an asset in fighting COVID?
- Has there been a viral mutation in Bangladesh? Is Bangladeshi COVID due to a different type of bat?
- Inherited from families.
- Acquired through hard living conditions such as poverty, labor, etc. Has the hot, humid weather helped fight COVID?
- Has the exposure to flu, dengue, chikungunya etc. bred cross-immunity?
- Even if they get infected, they can actively fight it off, [but how?].

Several of these points apply to India also, so we need to look for differences:

- Is Bangladeshi immunity enhanced by food, i.e., the lack of alcohol and the consumption of meat?
- Cleanliness e.g., practicing Muslims perform ablutions five times a day. Hindus may have similar prescriptions.
- Has communication of desired responses been better, and more effective in Bangladesh than in India?

Such questions could have been asked in March 2020. By then the mortality rate had been calculated at about 3.5%, i.e., it was an emergency, not a reason for national alarm. The Vice-Chancellor of East West University[8] forwarded a message from Stanford University on March 19, 2020. One point made therein was that 'This new virus is not heat-resistant and will be killed by a temperature of just 26/27 degrees Celsius (79-80 degrees Fahrenheit). It hates the Sun'. Already a good reason to suspect that the Bangladeshi poor, who face the heat of the midday sun regularly, would not be severely affected. These are the questions the intellectuals should have been curious about and asked. Of course, it is possible that existing conditions are not diagnosed properly, or improperly treated. But neither of these conditions alters the Public Health issue that is our primary concern.

Of course, it is not true that the working poor of Bangladesh are actually immune to COVID-19, but there is a possibility that the claim is true for practical purposes. In a study of three slums in Bangladesh, repeated four times so as to cover the entire period of the virus till the Omicron variant, we have found only 4 cases of severe COVID-19 in a study that questioned some 255 people directly and some 2500 people indirectly (see Basher et al. 2021). Hence our claim that the working poor of Bangladesh are practically immune to COVID. If the working poor of Bangladesh really do possess some form of immunity, then one has to ask why this question has not been more actively investigated. We have tried to publicize our results through a newspaper article, through the worldwide social sciences research network, by writing to researchers at major universities and organizations and actively through our own contacts. We may well be wrong, but it is surely a hypothesis based upon facts and whose importance requires urgent consideration. The larger question raised by our experience relates to the general unresponsiveness of the intellectual elite to questions of practical relevance. Since the problems we point to are general and not specific we will avoid giving names but each claim can be documented.

How then should we approach epidemics, both in the short and the long run? Fatality and debility rates are crucial. In the short run, there are two approaches. The 'loving' one and the 'clinical' one.

---

[8] Place of work of first two authors.

Names can be offensive, so some thought has been given to these names, which can be altered if a good reason is provided. We now describe each approach.

There is much to be said for considering each life to be infinitely valuable. Certainly no one would want to put a number on any one person's life. Public policy based on the assumption that each life is infinitely valuable can be called the loving approach. Yet this is not the approach to use when we consider public policy. We will emphasize this by making the point in two steps. We do not expect policymakers to make rules as though each life was infinitely valuable and we ourselves do not act as though our lives are infinitely valuable.

If you put up speed bumps every 50 feet and had three policemen at every road crossing it is quite conceivable that at least one life will be saved in Dhaka city. If each life is truly of infinite value why do we not do so? Why do we not agitate and march for such measures? Do not our actions show that we hold our politician's innocent at such times.

We often go from one market to another because of cheaper prices, say from Gulshan 1 to New Market. Perhaps you have done so too. Even if there is one chance in 10 million that you will get hurt in this short trip you should not have taken the trip. And yet everyday thousands of people make this extra trip. Clearly, they themselves do not put an infinite value on their own lives.

The loving approach is very admirable but it paralyzes all public policy. In dealing with the health of a nation, the more reasonable approach is to say that 'all lives are *equally* valuable'. This allows us to avoid discrimination, yet gives a principled approach to questions of epidemics and public health. We must give credit to the Government of Bangladesh for preaching the one and practicing the other. Given the silence of the intellectual class, it really had little choice.

## II. A Commendable Approach

The working hypothesis then is that for whatever reason Bangladeshi working people are minimally affected by COVID-19. This hypothesis was open to anyone to examine and critically evaluate. Why was this not done? There is something very static, immobile and self-assured about the intellectual elite which needs further discussion. It is these limitations of our mental approach to pandemics that requires critical rethinking.

The difference between the practical approach and a theoretical one is readily illustrated. Theoretically one will want to know who was affected, how they were infected, how they spread it to others, what are the effects on all parts of the human body, are there any long term impacts, do they have psychological repercussions and so on. It involves a hunt for fundamental causes and their modes of action. Practically the question is much simpler: what are the *effects* of COVID-19? If these effects are limited to the symptoms of the flu, or the common cough and cold that the Bangladeshi population suffer every year, then, for all practical purposes, COVID-19 *is* a cold. If it is not a cold, then who is it not a cold for? Are they distinguishable? Are they separable? By asking these questions, public policy can be based upon finding ways to distinguish and separate those particularly at risk and tending to them. There is no need whatsoever to understand the deeper biology and virology of the infection or its subsequent psychological effects. A failure to be clear about this fundamental issue has beset both the short and the long term approaches toward COVID-19.

If we look at the behavior of our doctors during the worst of the pandemic we find intelligent pragmatic and cooperative approaches to the problem. Without any central guidance or direction, doctors began trading their experiences with each other, sharing information about medicines that seem to work, as well as best practices for care. It was the experience on the ground that mattered to them. Of course, they could have made mistakes and some probably did but this purist critique misses the point: *in the midst of a pandemic even inaction is action*. Faced with the crisis and with no clear guidance, the doctors adopted the commendable approach of letting the facts dictate their actions. The reality is that they faced dictated their behavior and actions. This highly commendable approach is not what we have seen from those who observed the world from ivory towers. Why is there such a disconnect between those who solve problems and those who think about them?

Perhaps the mindset is best described as intellectual colonialism. Whatever was said by the CDC[9] in the United States, or by other reputed international institutions, was taken to be the operative truth for Bangladesh, period. This approach assumes that practically, in terms of actual day-to-day lives and actions, the biological status of all human beings is identical. Otherwise one does not know why results that are proved for Europe or North America necessarily apply to Bangladesh in the same form and intensity. For practical purposes, it is not the overall similarity, but the practical differences, that may have to guide action on the ground. Even the famous ICDDRB[10], world renowned for many medical innovations, has shown no interest or activity in this regard. One can therefore understand the silence of the other international organizations. It is not even an updated mindset. As Miguel and Mobarak (2021) point out, economists had provided guidelines on the trade-off between public health and the economy in early 2020. If this approach had been applied and refined, our conclusions would emerge.

One has to read between the lines in order to glean information about possibilities that the mainstream of our content to gloss over. Refined that the literature has provided instances of natural immunity, effectiveness of non-vaccine medicines, radiation in susceptibility to COVID-19, and correlation of COVID-19 with other biological features of Bangladeshis. Somehow this colonized mindset is unwilling to think of the human body as so complex an organism that a number of minor variants may not produce a major result.

---

[9] Centers for Disease Control and Prevention.
[10] International Centre for Diarrhoeal Disease Research, Bangladesh.

## III. Do Genes Matter?[11]

Science assumes that all COVID virus are identical and that all humans are identical. This is a convenient assumption as it allows data and theories worldwide to be assimilated into one enormous sample. But is it necessarily true? Are Bangladeshi bats, the presumed host of the virus, the same as Chinese bats? Are the microbiota of poor Bangladeshi identical with the microbiota of rich Bangladeshi? Let alone the microbiota of Western people? The initial panic caused by a genetic study showing that Bangladeshis were particularly susceptible to COVID-19 because they possessed an Neanderthal gene in exceptional proportions (Zeberg and Pääbo, 2020). The warning was quite understandable but the subsequent reaction is inexplicable. Of the many genes in the human body only one was identified as peculiar to Bangladesh. We do not know how this gene interacts with other features of the human body and whether there were not multiple effects of this gene which would counteract the effects of the Neanderthal gene. This in itself should have been a major item for research, yet we have seen no steps, at least in public, on this question. We find that there are significant statistical differences between the susceptibility to COVID-19 between the East and the West, that thalassemia is correlated with a lower incidence of COVID-19, and that natural immunity does seem to occur (El-Battrawy et al., 2021). Why can all these factors not act in conjunction and provide Bangladeshi workers with significant natural immunity to COVID? It is a question worth investigating.

Scientific journals create a further wall by the institution of 'ethical approval'. If the question is 'how do families cope when one parent goes abroad to work?', many delicate questions may arise and truthful answers can pry into a family's privacy. Immigration is a long term phenomenon, very little of consequence changes in three months. In this case, the safeguards of ethical clearance are justified. But the situation is entirely different in a pandemic. The entire population being surveyed could be radically changed through death and disability.  But what can be the privacy violations when subjects are told of the topic and consent to being asked? In its inability to recognize the demands of specific emergencies, scientific standards of ethical clearance are actually hurting the acquisition of knowledge when it matters. This is counterproductive.

The social scientists have been similarly blind, perhaps more so. The major question to be addressed has two components; first what are the immediate actions needed to alleviate suffering and prevent the spreading of the disease and secondly what steps should now be taken to prepare ourselves for a recurrence of such a viral disease. It is the task of the intellectuals to be most concerned with setting up a framework for the second question, even as they suggest practical steps to meet the immediate crisis. Where is this framework? Many pertinent questions have been raised about budgets, inflation, cash grants, help for SME's and so on. But why has there not been but deeper study of the potential of pandemics and the provision of guidelines for a permanent response? There is no reason to believe that while we are active in providing vaccines for the virus, the virus is yet busier finding mutations for our vaccines. It is a long haul. Satisfying ourselves with short run responses shows a failure to understand and appreciate the deeper problems that must arise in a globalized world.

---

[11] The approach of this section is missing from Miguel and Mubarak (2021).

## IV.  Ethical Approval or Unethical Hindrance?

The major result of our survey is that, for all practical purposes, Bangladeshi workers can be left free to work for their livelihood and continue with their lives because they appear immune to the *effects* of COVID. This is the question aroused by our findings[12] about Bangladeshi slum dwellers. It is natural to ask if we have tried to publicize the finding that has struck us as needing immediate research. Among those who have objected to our results, it has been said by the referee of a famous Bangladeshi economics journal that, if there are two brothers, one in the village and one in a slum, do they really have different susceptibilities to the virus? This misses the major point of the results: both the brothers may well have immunity, with the urban brother being subject to greater hazard of COVID due to life in the slum. Since the urban brother shows unusual resistance to COVID, hence the rural brother will be even more so. Upon examining the issue, nothing that the referee said weakens our arguments.

Even if the long run was the concern of the intellectuals, we cannot find any evidence of it on the ground. If the aim was to set up a gold standard of evidence about pandemics, then we should have had a concerted call for new courses, new programs, and new experimental studies of pandemics and their impact upon human societies. As far as we can tell from the media, there has been a complete blank on this issue. What is the directive principle: in practical matters, it is orders of magnitude that are decisive. Showing us some one paper presented at some one seminar and some one letter being written to some one newspaper is a curiosity. It is totally out of proportion with the magnitude of the problem to be solved and is almost laughable. When we say the intellectual reaction has been utterly inadequate, we mean only one thing—that the reaction has been quite out of proportion to the magnitude of the problem.

Foreign academics are always interested in Bangladeshi problems. Why then has this research not been published in a foreign journal with all its details? The main problem here is an interesting one: it concerns the previously mentioned 'ethical approval'. This is a very well-meant safeguard to prevent vulnerable people from being exploited by researchers. However, it is strictly applicable only to cases

---

[12] See Basher et al. (2021).

where detailed personal questions are being asked or identities are being revealed. It is of passing relevance for a study which questions individuals about the presence of illness in their households after taking their permission. It is especially hard to impose the ethical approval requirement in the midst of a pandemic when the need for urgent and accurate knowledge is acute. We began the study in March 2020, at a time when there was a general alarm, bordering on panic. Universities were closed and it was difficult to get into contact with any authorities. Even those who do provide such ethical approval do so by committee and getting all the committee members together, having them agree on a meeting, sending their comments and then revising the questions would have taken a few months at the earliest. We decided that it was more important to begin the process and then worry about formalities, mistakenly as we subsequently found out. No reputable foreign journal will consider a paper that has not approved *prior* ethical approval, even though the study is as simple and straightforward as that of this paper, and when all the respondents are asked beforehand if they agree to the study. In short, foreign academics need to adapt to the realities of a pandemic and the urgency of practical knowledge.

## V. Sequencing the Future

If only 4% of the population are affected by the disease, in the sense of debilitation or fatality, then this is really a problem of public health. We cannot have a country stand still and jeopardize the economic and social future of the 96% for the sake of the 4%. Think of all the hospitals doctors and nurses and public health staff and activities that could have been funded if the lockdowns had been avoided and that money used towards mitigating the effects of COVID, instead of a somewhat desperate effort to suppress it.

To repeat: While we busily develop vaccines for the virus, the virus is busier creating mutations for our vaccines. Who will win? No one knows. What is incumbent upon us is the extraction of all the valuable knowledge and practice from the past so that the future finds us better prepared.

**Declarations**

- **Funding:** The research was entirely done with private (authors' own) funds.

- **Competing interests:** We declare that we have no conflict of interest.

- **Ethics approval and consent to participate:** Ethical guidelines were followed and we asked for permission from respondents before asking questions.

- **Trial registration details:** Not applicable.